\documentclass[twocolumn,amsmath,amssymb]{revtex4}

\usepackage{graphicx}
\usepackage{dcolumn}
\usepackage{bm}
\newcommand{\etal}{\emph{et al.}}

\begin{document}      

\title{Charge-ordering, commensurability and metallicity in the phase diagram of layered 
$\bf\rm Na_xCoO_2$.
} 
\author{Maw Lin Foo$^{1}$, Yayu Wang$^{2}$, Satoshi Watauchi$^{1\dagger}$, H. W. 
Zandbergen$^{3,4}$, Tao He$^{5}$, 
R. J. Cava$^{1,3}$, and N. P. Ong$^{2,3}$
}

\affiliation{
$^1$Department of Chemistry, $^2$Department of Physics, $^3$Princeton Materials 
Institute, Princeton University, New Jersey 08544, U.S.A.\\
$^4$National Centre for HREM, Laboratory of Materials Science, Delft University of 
Technology, The Netherlands \\
$^5$DuPont Central Research and Development Experimental Station,  Wilmington,  Delaware 
19880, U.S.A.
}

\date{\today}      

\begin{abstract}
The phase diagram of non-hydrated  $\rm Na_xCoO_2$ has been determined by changing the 
Na content $x$ using a series of chemical reactions.   As $x$ increases from 0.3, the 
ground state goes from a paramagnetic metal to a charge-ordered insulator (at 
$x=\frac12$) to a `Curie-Weiss metal' (around 0.70), and finally to a weak-moment 
magnetically ordered state ($x>0.75$).  The unusual properties of the state at $\frac12$ 
(including particle-hole symmetry at low $T$ and enhanced thermal conductivity) are 
described.  The strong coupling between the Na ions and the holes is emphasized.  
\end{abstract}
\maketitle                   
Research on oxide conductors has uncovered many interesting electronic states 
characterized by strong interaction, which include unconventional superconductivity, and 
charge- or spin-ordered states~\cite{Tokura,Maeno}.  Recently, attention has focussed on 
the layered cobaltate $\rm Na_xCoO_2$.  At the doping $x\sim\frac23$, $\rm Na_xCoO_2$ 
exhibits an unusually large thermopower~\cite{Terasaki}.  Although the resistivity is 
metallic, the magnetic susceptibility displays a surprising Curie-Weiss 
profile~\cite{Ray}, with a magnitude consistent with antiferromagnetically coupled 
spin-$\frac12$ local moments equal in number to the hole carriers~\cite{Wang}.  The 
thermopower at 2.5 K is observed to be suppressed by an in-plane magnetic 
field~\cite{Wang}.  This implies that the enhanced thermopower is largely due to spin 
entropy carried by strongly correlated holes (Co$^{4+}$ sites) hopping on the triangular 
lattice.  When intercalated with water, $\rm Na_xCoO_2$$\cdot y$$\rm H_2O$ becomes 
superconducting at or below 4 K~\cite{Takada} for 
$\frac14<x<\frac13$~\cite{Cava,MIT,Jin}.  These experiments raise many questions.  Is 
the Curie-Weiss state at $\frac23$ continuous with the $\frac13$ state surrounding 
superconductivity?  Are commensurability and charge-ordering effects important?  To 
address these questions, we have completed a study of the phase diagram of non-hydrated 
$\rm Na_xCoO_2$.  As $x$ increases from 0.3 to 0.75, we observe a series of electronic 
states, the most interesting of which is an insulating state at $x = \frac12$ that 
involves charge ordering of the holes together with the Na ions.  We identify details 
specific to the triangular lattice, especially in the metallic state from which the 
superconducting composition evolves, and comment on recent theories.

\begin{figure}[h]                       
\includegraphics[width=7cm]{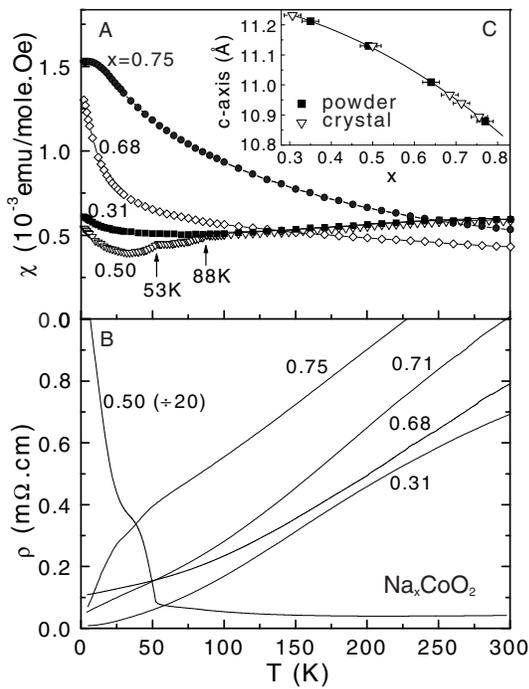}
\caption{\label{chi} The susceptibility $\chi$ (A) and in-plane resistivity $\rho$ (B) 
of single crystals of $\rm Na_xCoO_2$ with $x$ determined by ICP (C).  In Panel A, 
$\chi$ is  measured in an in-plane field $H$ = 5 T ($\bf H\perp \hat{c}$).  In the 
crystal with $x$ = 0.75, $\chi$ fits the Curie-Weiss form $\chi = C/(T+\theta)$ (with 
$\theta\sim$ 150 K and $C$ = 2.81 cm$^3$K/mole).  For $x$ = 0.5, sharp transitions are 
observed at $T_{c1}$ and $T_{c2}$ (arrows).  Panel B shows the $T$ dependence of $\rho$ 
at selected $x$.  Insulating behavior is observed at $x$ = 0.5 (data displayed at lower 
scale) in contrast to metallic behavior in the rest.  At low $T$, $\rho$ is $T$-linear 
for $x$ = 0.71 but varies as $T^2$ for $x$ = 0.3.  In Panel C, the $c$-axis lattice 
parameter measured by XRD is plotted against the Na content $x$ fixed by ICP in powder 
samples.   } 
\end{figure}

Starting with powder or single-crystal samples with $x\sim$ 0.75, we vary $x$ by 
specific chemical deintercalation of Na (Fig. \ref{chi}, caption).   Powders of 
Na$_{0.77}$CoO$_2$ were made by solid-state reaction of stoichiometric amounts of 
Na$_2$CO$_3$ and Co$_3$O$_4$ in oxygen at 800 C.  Sodium de-intercalation was then 
carried out by treatment of samples in solutions obtained by dissolving I$_2$ (0.2 M, 
0.04 M) or Br$_2$ (1.0 M) in acetonitrile. After magnetic stirring for five days at 
ambient temperature, they were washed with copious amounts of acetonitrile and multiple 
samples were tested by the ICP-AES method to determine Na content. Unit-cell parameters 
were determined by powder X-ray diffraction (XRD) with internal Si standards. For the 
transport studies, we first grew a boule (with $x$ = 0.75) in an optical furnace by the 
floating-zone technique.  Crystals cleaved from the boule (1.5 mm $\times$ 2 mm $\times$ 
0.2 mm) were immersed in solutions of I$_2$, Br$_2$ in acetonitrile or NaClO$_3$ in 
water for up to 2 weeks to deintercalate sodium.  Because of the strong sensitivity of 
the electronic states to $\Delta x$, it was essential to fine-tune the de-intercalation 
times and conditions until high uniformity of the Na distribution was achieved (as 
measured by the line-width of the XRD).   Moreover, by combining powder XRD and 
inductively-coupled plasma analysis (ICP) experiments, we measured accurately the 
variation of the $c$-axis lattice parameter vs. $x$ (Fig. \ref{chi}C).
Figure \ref{chi}A shows the systematic variation of the magnetic susceptibility $\chi$ 
with $x$.  Previous work~\cite{Ray,Wang} shows that for $x \sim \frac23$, $\chi$ vs. $T$ 
follows the Curie-Weiss law $\chi = C/(T+\theta)$ with $\theta\sim$ 70 K.  Above 
$\frac23$, we find that, at $x$ = 0.75,  $\chi$ is slightly rounded below 20 K, 
consistent with the appearance of a weak magnetization $M$ ($\sim 0.03\; \mu_B$ per Co 
with $\mu_B$ the Bohr magneton).  A spin-density-wave (SDW) state at 0.75 has been 
suggested~\cite{Motohashi,Sugiyama}.  When we decrease $x$ below $\frac23$, the 
Curie-Weiss divergence in $\chi$ is progressively reduced with decreasing $x$ until it 
vanishes at $\frac12$.   Close inspection of $\chi$ at $\frac12$ reveals the existence 
of 2 sharp cusps at $T_{c1}$ = 88 K and $T_{c2}$ = 53 K (arrows).  These 2 transitions 
signal the onset of an insulating state which we discuss shortly.  In this 
charge-ordered state, $\chi$ remains independent of the magnetic field $H$, which is 
inconsistent with magnetic ordering.  With further decrease of $x$ below $\frac12$ to 
$x$ = 0.31, the profile of $\chi$ becomes relatively featureless, although its magnitude 
remains large compared with the Pauli susceptibility in conventional metals.  

%
%
%
\begin{figure}[h]                       
\includegraphics[width=7cm]{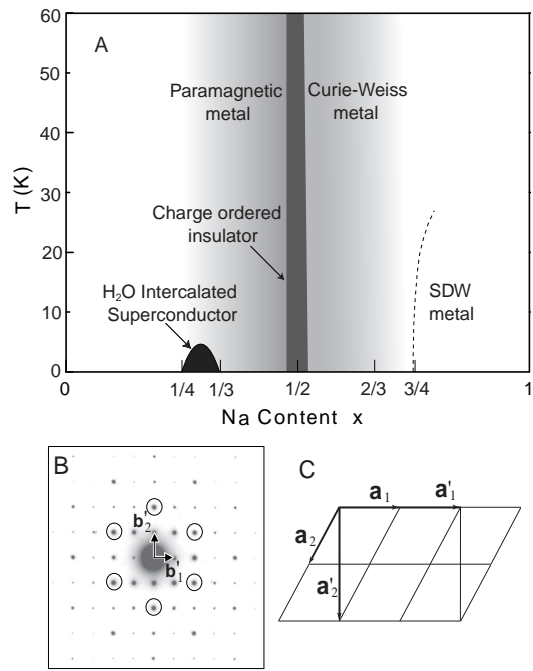}
\caption{\label{phase} The phase diagram of non-hydrated $\rm Na_xCoO_2$ (A), electron 
diffraction pattern of a crystal with $x=\frac12$ (B) and the Na superstructure lattice 
(C).  In Panel A, the charge-ordered insulating state at $\frac12$ is sandwiched between 
the paramagnetic metal at 0.3 and the Curie-Weiss metallic state at 0.65-0.75.  The 
superconducting state is obtained on intercalation with H$_2$O~\cite{Takada,Cava}.    
In Panel B, the diffraction pattern is taken along [001] with 200 kV electrons in a 
Phillips CM200 electron microscope.  Bragg spots of the hexagonal lattice are encircled.  
The spots located at $n_1{\bf b}'_1+n_2{\bf b}'_2$ correspond to the supercell with 
lattice vectors ${\bf a}'_1$ and ${\bf a}'_2$ drawn in Panel C.
}
\end{figure}
In contrast with the smooth evolution of $\chi$ vs. $x$, the in-plane resistivity $\rho$ 
shows a dramatic change in behavior across $x = \frac12$ (Fig. \ref{chi}B).  At all $x$ 
values away from $x = \frac12$, $\rho$ is `metallic' in its $T$ dependence.  For e.g., 
in the well-studied Curie-Weiss state ($x$ = 0.68, 0.71), $\rho$ has a characteristic 
$T$-linear profile below 100 K~\cite{Wang}.  At higher doping ($x$ = 0.75), $\rho$ shows 
a distinct change-in-slope near 20 K in agreement with previous work~\cite{Sugiyama}.  
This reflects the onset of weak magnetic order discussed above.  In contrast to these 
metallic states, insulating behaviour abruptly appears at $x= \frac12$.  Initially, 
$\rho$ in this sample increases gradually as $T$ falls towards $T_{c1}$ and $T_{c2}$, 
where it exhibits weak anomalies (note change in scale).  Below $T_{c2}$, however, 
$\rho$ rises rapidly to reach $\sim 20$ m$\Omega$cm at 4 K.  This insulating state is 
confined to a narrow interval in $x$.  At our lowest doping $x$ = 0.31, we recover a 
high-conductivity metallic state.  Below 30 K, $\rho$ follows a $T^2$ behavior and falls 
to a value (9 $\mu\Omega$cm) that is $\sim$5 times lower than at $x = 0.71$.    

The results from $\chi$ and $\rho$ imply the phase diagram displayed in Fig. 
\ref{phase}.  The dominant feature in the $T$-$x$ plane is the narrow insulating state 
at $x = \frac12$ which separates two distinct metallic states.  Below $\frac12$, we have 
a `paramagnetic metal' with high conductivity whereas, above $\frac12$, we have a 
`Curie-Weiss' metallic state in which a $T$-linear resistivity~\cite{Wang} coexists with 
a $\chi$ that is Curie-Weiss like~\cite{Ray,Wang}.   

What is the nature of the insulating state at $x = \frac12$?  An important feature 
specific to this state is revealed by electron-diffraction studies (Fig. \ref{phase}B).  
Even at 300 K, the Na ions at $x = \frac12$ order as a superstructure with lattice 
vectors $a\sqrt{3}\; {\bf \hat{x}}$ and $2a\;{\bf \hat{y}}$ in the basal plane with $a$ 
the hexagonal lattice parameter.  By contrast, at other doping levels, the 
superstructure Bragg spots are either much weaker or absent.   

%

%
\begin{figure}[h]                       
\includegraphics[width=8cm]{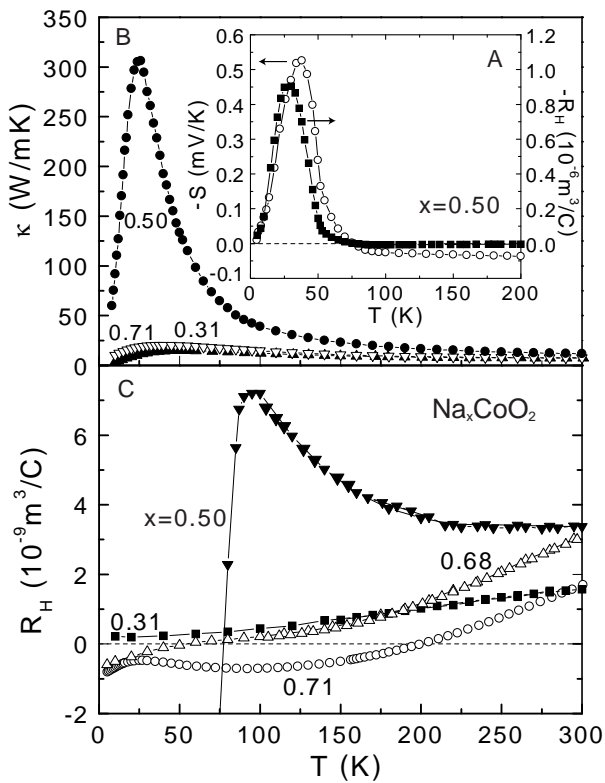}
\caption{\label{kappa} The thermopower $S$ and the Hall coefficient $R_H$ in $\rm 
Na_xCoO_2$ for $x = \frac12$ (A), and plots of the in-plane thermal conductivity 
$\kappa$ (B), and $R_H$ (C) for several values of $x$.  In Panel A, both $|S|$ and 
$|R_H|$ begin to increase steeply just below $T_{c2}$ (88 K) and attain maxima at 40 and 
30 K, respectively (note $-S$ and $-R_H$ are plotted).  Below the peak values, they fall 
towards zero as $T\rightarrow 0$, consistent with a particle-hole symmetric state.  In 
Panel B, $\kappa$ at $x = \frac12$ rises to a very large peak value comparable to that 
of quartz, whereas it remains small at all other $x$.  Panel C shows that the curve of 
$R_H$ vs. $T$ at $x= 0.50$ is qualitatively different from all other $x$ values.  In the 
metallic samples, $R_H$ increases with $T$, consistent with strongly interacting 
carriers hopping on a triangular lattice.
}
\end{figure}

The long-range nature of the Na superstructure is brought out clearly by the thermal 
conductivity $\kappa$ measured parallel to the layers (as in layered oxides, $\kappa$ 
here is dominated by the phonons).  Because phonons are strongly scattered by disorder 
in the Na sublattice, we expect the phonon mean-free-path $\ell_{ph}$ to be much longer 
at $\frac12$ than at neighboring $x$ values.  This is strikingly confirmed by the curves 
of $\kappa$ vs. $T$ (Fig. \ref{kappa}).  In the metallic samples ($x$ = 0.31 and 0.71), 
$\kappa$ is only weakly $T$ dependent and relatively small at all $T$.  However, in the 
crystal with $x= \frac12$, it rises steeply below $T_{c2}$ to 300 W/mK, an exceptionally 
high value that is comparable to that observed in high-purity quartz and diamond.  Hence 
both the diffraction and $\kappa$ results provide strong evidence that, in sharp 
contrast with the metallic states, the Na ions at $x = \frac12$ order with very long 
range correlation even at $T\sim$ 300 K.

Further insight into the insulating state is provided by the Hall coefficient $R_H$ 
(Figs. \ref{kappa}A and C).   At $\frac12$, we find that, above 200 K, $R_H$ is 
hole-like with a $T$-independent value equivalent to a Hall density $1/eR_H$ = 
1.8$\times 10^{21}\;\mathrm{cm}^{-3}$.  (In metallic samples, however, $R_H$ is $T$ 
dependent; see below).  Near $T_{c1}$, $R_H$ rises to a weak maximum and then plunges to 
large negative values.  Below $T_{c2}$, it increases dramatically in magnitude to a peak 
value 250 times greater than its magnitude above 200 K.  The giant increase in $|R_H|$ 
implies that the itinerant carrier density $n$ decreases by $\sim$2 orders of magnitude.  
The profile of the in-plane thermopower $S$ is qualitatively similar.  $S$ changes sign 
near $T_{c1}$, and attains a large negative peak below $T_{c2}$.  However, the relative 
increase in magnitude ($\sim$20) is smaller.  

The behavior of $R_H$ at low $T$ reveals an unusual feature of the insulating state.  
$|R_H|$ falls steeply towards zero as $T\rightarrow 0$.  This contrasts with the usual 
behavior in semiconductors where it diverges as $1/n$ (as $n\rightarrow 0$).  In 
general, we have $R_H = \sigma_H/H\sigma^2$, with $\sigma_H$ the Hall conductivity and 
$\sigma= 1/\rho$ the conductivity.  The decrease of $R_H$ to zero as $T\rightarrow 0$ 
implies that $\sigma_H$ must vanish faster than the rate at which $\rho^2$ is 
increasing.  The vanishing of the Hall current at low $T$ implies that the charge 
excitations exhibit strict particle-hole symmetry.  They are neither hole-like nor 
electron-like in their response to the applied field.  

Previous transport~\cite{Wang,WangHall}, magnetic~\cite{Ray}, and angle-resolved 
photoemission (ARPES)~\cite{Hasan} experiments point to the conclusion that strong 
correlation is essential for understanding the electronic properties of $\rm Na_xCoO_2$.  
On the triangular lattice in the CoO$_2$ layer, a fraction $(1-x)$ of the sites are 
occupied by itinerant spin-$\frac12$ holes (Co$^{4+}$) hopping in a diamagnetic 
background of Co$^{3+}$ ions~\cite{Wang}.  The hopping amplitude $t$ for $x\sim \frac23$ 
measured by ARPES~\cite{Hasan} is quite small ($|t|\sim$ 7 meV).  The Fermi Surface 
encircles the $\Gamma$ point, consistent with a negative sign for $t$.

Recently, several groups have applied the $tJ$ Hamiltonian to $\rm 
Na_xCoO_2$~\cite{Baskaran,Shastry,Lee,Motrunich}.  The derived phase diagram is strongly 
informed by the classical lattice-gas model which predicts that, on the triangular 
lattice, the electrons (holes) crystallize into a $\sqrt{3}\times\sqrt{3}$ structure at 
$x = \frac13$ ($\frac23$)~\cite{Motrunich}.  However, at $\frac12$, commensurability 
effects are weakest and charge ordering is not favored.  We also note that the density 
of states calculated in 2D tight-binding approximation~\cite{Shastry,Lee,Baskaran} 
reveals no sharp anomaly at $x = \frac12$ if $t<0$ (a van-Hove singularity does occurs 
at $x = \frac12$ if $t>0$).  Hence the observed insulating state is quite unexpected.  

The phase diagram in Fig. \ref{phase} reveals a characteristic feature of $\rm 
Na_xCoO_2$, namely, the strong influence of the Na ions on the electronic properties.  
The occurence of the insulating state seems to require a strong interaction between the 
ions and holes even though they occupy separate layers.  For either charge species alone 
on the triangular lattice at $x = \frac12$, the liquid state seems to be more stable at 
finite temperature.   However, if both populations are present, density fluctuations in 
one will strongly influence the other.  The incipient localization of the holes is 
evident at 300 K as indicated by the weakly rising $\rho$ (Fig. \ref{chi}B).  In turn, 
as the amplitude of the hole-density modulation grows, it becomes more favorable 
energetically for the Na ions to order.  The strong coupling leads to a steep increase 
in both modulation amplitudes until a transition occurs in which the carriers condense 
into an insulating charge-ordered state.  First-principles density functional 
calculations for $x = \frac12$~\cite{Singh} show a very sharp peak in the density of 
states at the Fermi energy.  This may account for the enhanced sensitvity to Na ordering 
at this particular doping.

It is interesting to compare the paramagnetic metal with the Curie-Weiss metal.  As 
noted, at low $T$, $\sigma$ is $\sim$5 times higher in the former.  In the sample with 
$x =$ 0.31, the nearly $T$-independent value of $R_H$ at low $T$ gives a Hall density = 
$2.8\times 10^{22}\mathrm{cm^{-3}}$, which is in agreement with the 2D hole density 
calculated from $n_{2D} = (2/\sqrt{3}a^2)(1-x)$, where $a$ = 2.82 \AA  is the lattice 
parameter of the triangular lattice.  The large hole density and high conductivity at 
low $T$ implies that screening of charge fluctuations is quite effective at 0.31.  
Accordingly, its transport properties are more conventional than those of the 
Curie-Weiss metal.  In the latter, the thermopower is greatly enhanced by a large 
spin-entropy component~\cite{Wang}.  The susceptibility has a Curie-Weiss profile with a 
magnitude consistent with a spin-$\frac12$ local-moment population equal to the hole 
concentration $(1-x)$.  These properties are anomalous from the viewpoint of a 
conventional metal.  Finally, as shown in Fig. \ref{kappa}C, $R_H$ is initially negative 
at low $T$, but becomes positive near 200 K (for $x$ = 0.71) and displays an unusual 
$T$-linear increase over a broad range of $T$~\cite{WangHall}.  Holstein~\cite{Holstein} 
has shown that coherent interference leads to a $T$-linear dependence of $R_H$ for 
electrons hopping on a triangular lattice.  A similar $T$-linear behavior has been 
derived for correlated holes by Shastry and coworkers~\cite{Shastry} for the 
high-frequency $R_H(\omega)$.  This anomalous behavior appears to be intrinsic to the 
triangular lattice in the regime $k_BT\ge |t|$.  The hole-like sign of $R_H$ at high $T$ 
is consistent with $t<0$ consistent with ARPES  ~\cite{Hasan,Motrunich,Shastry}.  The 
$T$-linear dependence of $R_H$ is present at $x = 0.31$ but much weaker.  As discussed 
above, the transport properties in these 2 metallic states are again very different from 
those in the charge-ordered phase at $x = \frac12$ that separates them.

Finally, we remark that the superconducting state produced by intercalation with H$_2$O 
evolves from the paramagnetic metal which is separated from the Curie-Weiss state by the 
charge-ordered insulator.  It is unlikely that antiferromagnetic correlations, so 
strongly visible at $x\sim \frac23$, play a role in the superconducting state.   
Moreover, in the paramagnetic metal, we have not observed any evidence for 
$\sqrt{3}\times\sqrt{3}$ charge ordering, nor any evidence for a ferromagnetic 
instability.  The large carrier density ($\sim 3\times 10^{22}\mathrm{cm^{-3}}$ at $x$ = 
0.31) is 10 times larger than in the cuprates (doped at 0.3).  However, the larger 
effective mass $m^*$ here may reduce the superfluid density $n_s/m^*$ to a value 
comparable to that observed in the overdoped limit in cuprates.  

We thank Thomas S. Connell for performing the ICP-AES analysis, and P. A. Lee for 
valuable comments.  The research at Princeton is supported by a MRSEC grant from the 
National Science Foundation (DMR 0213706).  MLF and RJC were partially supported by NSF 
Grant DMR 0244254.
\vspace{3mm}\\
$\dagger$\emph{Permanent address}: Center for Crystal Science and Technology, University 
of Yamanashi,
7 Miyamae, Kofu, Yamanashi 400-8511, Japan.

\end{document}